%% file: main.tex
\documentclass[10pt,twocolumn,letterpaper]{article}

\usepackage{wacv}              % To produce the CAMERA-READY version
\usepackage{amsmath}


\definecolor{wacvblue}{rgb}{0.21,0.49,0.74}
\usepackage[pagebackref,breaklinks,colorlinks,allcolors=wacvblue]{hyperref}
\usepackage{multirow}
\usepackage{graphicx}
\usepackage{subcaption}

\def\wacvPaperID{*****} % *** Enter the WACV Paper ID here
\def\confName{WACV}
\def\confYear{2027}

\title{Intellectual Property Protection of Hyperdimensional Computing Classifiers: Black-Box Ownership Verification via Model Watermarking}
\author{
Hamed Poursiami$^{*}$ \quad
Farshad Soleiman$^{*}$ \quad
Ahmad Tahmasivand$^{*}$ \quad
Guojing Cong$^{\dagger}$ \\
Thomas Potok$^{\dagger}$ \quad
Khaled N. Khasawneh$^{*}$ \quad
Kai Zeng$^{*}$ \quad
Maryam Parsa$^{*}$ \\[0.5em]
$^{1}$Electrical and Computer Engineering Department, George Mason University, Fairfax, VA, USA \\
$^{2}$Oak Ridge National Laboratory, Oak Ridge, TN, USA \\[0.3em]
{\small\tt{hpoursia@gmu.edu, fsoleima@gmu.edu, atahmasi@gmu.edu, congg@ornl.gov,}}\\
{\small\tt{potokte@ornl.gov, kkhasawn@gmu.edu, kzeng2@gmu.edu, mparsa@gmu.edu}}
}

\begin{document}
\maketitle
\input{Sections/abs}    
\input{Sections/intro}
\input{Sections/hdc}

\input{Sections/method}
\input{Sections/ExpSet}

\input{Sections/Results}

\input{Sections/hardware}

\input{Sections/conclusion}

{
    \small
    \bibliographystyle{ieeenat_fullname}
    \bibliography{main}
}

\end{document}

%% file: Sections/abs.tex
\begin{abstract}
Hyperdimensional computing (HDC) offers efficient learning and inference for resource-constrained systems, yet ownership verification of its models remains underexplored. We present the first systematic investigation of model watermarking in HDC by adapting backdoor-based embedding to three classifiers with different prototype-learning mechanisms. Across four benchmarks, with conventional neural networks as references, we examine whether watermarks can be learned without compromising task accuracy and whether they survive subsequent model modification and replication. The error-driven classifiers, AdaptHD and OnlineHD, support verification at a nominal trigger exposure of $1\%$ with at most $0.58$ percentage points of mean accuracy loss, whereas single-pass Vanilla HDC requires greater exposure and can incur substantial accuracy costs. The error-driven models retain verification under native continued training, but extraction produces surrogates that reproduce much of the task behavior without verifiable watermarks. Watermark overwriting readily embeds attacker watermarks into HDC prototypes and erodes the owner’s watermark more strongly than in the neural network references, although the strongest erosion coincides with large accuracy losses. We further deploy watermarked AdaptHD on an FPGA and stress its execution through undervolting and overclocking. Verification survives substantial computation errors, whereas more severe stress either causes verification failure together with accuracy collapse or results in inference timeouts. These findings establish the feasibility of watermarking in HDC while identifying limitations in retention under extraction and sensitivity to overwriting.
\end{abstract}

%% file: Sections/intro.tex
\section{Introduction}
\label{sec:intro}

Hyperdimensional Computing (HDC) has emerged as a compelling brain-inspired approach to efficient machine learning, with applications ranging from image and biosignal classification to robotic mapping and efficient decision making~\cite{ge2020classification,rahimi2016hyperdimensional,mitrokhin2019learning,snyder2026brain,poursiami2025scalable}. 
By representing information in high-dimensional vectors and processing it through lightweight vector operations, HDC supports rapid training, energy-efficient computation, and robustness to noise. Its computational efficiency and suitability for hardware acceleration on FPGAs and in-memory computing platforms make it particularly attractive for resource-constrained edge devices~\cite{salamat2019f5hd,karunaratne2020memory}. However, deployment outside the developer’s trusted infrastructure can expose stored model representations to unauthorized access.
In prototype-based HDC classifiers, these representations include class hypervectors that consolidate task-specific knowledge acquired from training data. Access to these prototypes and the corresponding encoder can enable replication of the classifier without repeating the investment in data acquisition, representation design, training, and optimization. Protecting the intellectual property (IP) embodied in these models is therefore an important requirement for their deployment.

Despite this need, IP protection for HDC models remains underexplored. StealHD~\cite{ma2026stealing} investigates data-free extraction through prediction access, highlighting the threat of unauthorized model replication. Existing protection approaches address different aspects of model confidentiality: HDLock~\cite{duan2022hdlock} uses privileged encoding to impede reverse engineering, while EP-HDC~\cite{park2025ephdc} protects class hypervectors through homomorphic encryption while enabling client-side inference. These mechanisms restrict access to model internals but do not themselves provide evidence linking a suspect copy to its original developer. Model watermarking addresses this complementary ownership-verification objective by embedding a secret signature that can later be tested in a suspect model~\cite{uchida2017embedding,soleiman2024randomized,adi2018turning}. Although extensively studied in deep neural networks~\cite{zhang2018protecting,lukas2022sok,xue2022intellectual} and investigated in spiking neural networks~\cite{poursiami2024watermarking}, its applicability to HDC classifiers remains an open research question.

HDC’s distinctive representation and learning mechanisms raise fundamental questions about how watermarks can be embedded and retained. In conventional HDC classifiers, training combines the hypervector representations of samples from each class into a shared prototype through superposition~\cite{kanerva2009hyperdimensional}. This associative structure provides a natural starting point for investigating whether a secret watermark can be incorporated alongside the information required for the primary task. However, the watermark and task-related information share the same prototypes, creating the potential for interference during embedding and changes to the watermark during subsequent learning. The central question is therefore how these learning mechanisms affect the ability to embed a verifiable watermark while preserving classification accuracy, and whether the resulting ownership evidence remains reliable after model modification or extraction.

To address these questions, we adapt the backdoor-based watermarking framework of Adi et al.~\cite{adi2018turning} to HDC and evaluate three representative classifiers: Vanilla HDC, AdaptHD, and OnlineHD. This approach provides a common basis for examining how these classifiers learn secret trigger–label associations and retain them after deployment. Our evaluation combines statistical verification, exposure–accuracy analysis, and tests of model adaptation, extraction, and competing watermark insertion, with conventional neural networks as references. We further deploy watermarked AdaptHD models on an FPGA to examine verification under hardware execution faults. Together, these experiments assess both embedding feasibility and the survival of ownership evidence under post-deployment threats and hardware faults. Our contributions are:
\begin{itemize}
\item \textbf{A systematic investigation of watermarking in HDC.}
To the best of our knowledge, we present the first study of model watermarking in HDC, adapting backdoor-based embedding to three prototype-learning mechanisms and assessing verification through a binomial test with empirical null calibration.

\item \textbf{Characterization of embedding feasibility and post-deployment limits.}
At 1\% trigger exposure, AdaptHD and OnlineHD verify across all benchmarks with at most 0.58~pp mean accuracy loss, whereas Vanilla HDC suffers from superposition dilution. We show that embedded watermarks survive white-box adaptation and cannot be overwritten by competing triggers without either leaving the owner's signature verifiable or collapsing task accuracy (10.37 -- 42.21~pp drop), whereas black-box extraction removes OOD watermarks across all model families.

\item \textbf{FPGA validation under hardware faults.}
We deploy watermarked AdaptHD inference on a ZCU104 FPGA and evaluate its response to undervolting and overclocking. Verification survives operating conditions that reduce classification accuracy by up to $22.00$ pp, while more severe stress causes verification failure or inference timeouts.

\end{itemize}

%% file: Sections/hdc.tex
\section{Preliminaries: HDC Classification}

Hyperdimensional Computing, also known as Vector Symbolic Architecture (VSA), represents and processes information in a high-dimensional space $\mathcal{H}$ of dimensionality $D$ (typically $D \sim 10^3\text{–}10^4$) using distributed representations called hypervectors~\cite{kanerva2009hyperdimensional,kleyko2022survey1}.
Because independently sampled random vectors in high-dimensional space are nearly orthogonal with high probability, HDC can compose and manipulate symbolic and numerical structures using three fundamental algebraic operations:  (1) \textbf{Bundling} (superposition), which aggregates hypervectors into an associative composite preserving similarity to its inputs via element-wise addition in bipolar/real spaces or majority voting in binary spaces; (2) \textbf{Binding} (association), which combines two hypervectors into a dissimilar (quasi-orthogonal) hypervector via Hadamard product ($\otimes$) in bipolar/real spaces or XOR in binary spaces; and (3) \textbf{Permutation} (ordering), which applies a deterministic cyclic coordinate shift to encode positional or structural hierarchy.

\subsection{Encoding}

Let $\mathcal{X} \subseteq \mathbb{R}^F$ denote an $F$-dimensional input space and $\mathcal{Y} = \{1, \dots, C\}$ denote the set of $C$ target classes. An HDC classifier first maps an input $x = (x_1, \dots, x_F) \in \mathcal{X}$ into a $D$-dimensional hypervector $h = \phi(x)$ through an encoding module. For general feature-vector inputs, two foundational encoding families are widely used:
\begin{itemize}
    \item \textit{Record-based (or ID-level) encoding}~\cite{rahimi2016robust,imani2017voicehd}, which binds randomly generated feature-position hypervectors $\mathbf{p}_j$ with quantized feature-level hypervectors $\mathbf{l}(x_j)$ and bundles them across features ($h = \sum_{j=1}^F \mathbf{p}_j \otimes \mathbf{l}(x_j)$).
    \item \textit{Nonlinear random projection (sinusoid) encoding}~\cite{imani2019framework,rahimi2007random}, which projects continuous inputs via $W \in \mathbb{R}^{D \times F}$ ($W_{ij} \stackrel{\text{i.i.d.}}{\sim} \mathcal{N}(0,1)$) and phase shifts $b \in \mathbb{R}^D$ ($b_d \stackrel{\text{i.i.d.}}{\sim} \text{Uniform}[0, 2\pi]$) as $h = \cos(xW^\top + b) \otimes \sin(xW^\top)$ to approximate shift-invariant kernel similarity.
\end{itemize}
Beyond these general-purpose vector encodings, specialized HDC encoders have been designed for specific domains and modalities, including graph encodings for relational structures \cite{cong2026hypergraphx,poursiami2026vs,poduval2022graphd}, spatiotemporal $n$-gram encodings for text and multichannel biosignals ~\cite{joshi2016language,rahimi2016hyperdimensional}, resonator and fractional power encodings for spatial visual scenes~\cite{frady2022computing,renner2024neuromorphic}, and event-based encodings for neuromorphic dynamic vision sensors ~\cite{mitrokhin2019learning,zou2022eventhd}. 
\subsection{Associative Inference}
Classification is performed using an associative memory storing $C$ trainable class prototypes $\mathbf{P} = \{\mathbf{c}_1, \dots, \mathbf{c}_C\}$, where each prototype $\mathbf{c}_y \in \mathbb{R}^D$ consolidates the encoded patterns of class $y \in \mathcal{Y}$. Given a query input $x$ with encoded hypervector $h = \phi(x)$, the model predicts the class whose prototype maximizes cosine similarity:
\begin{equation} 
   \hat{y}(x) = \arg\max_{y \in \mathcal{Y}} \delta(h, \mathbf{c}_y), \qquad \delta(u, v) = \frac{\langle u, v \rangle}{\Vert{}u\Vert{}_2 \, \Vert{}v\Vert{}_2},
\end{equation}
\subsection{Prototype Learning Mechanisms}
Given a labeled training set $\mathcal{D} = \{(x_i, y_i)\}_{i=1}^N$, HDC classifiers construct and refine the class prototypes $\mathbf{P}$ on top of the encoder $\phi$. Among the prototype learning schemes developed for HDC, three widely used paradigms illustrate the progression from single-pass aggregation to iterative error-driven refinement:
\begin{itemize}
    \item \textbf{Vanilla HDC}~\cite{kanerva2009hyperdimensional,rahimi2016robust}: Each class prototype is constructed in a single pass by directly summing the encoded hypervectors of all training samples belonging to that class:
        \begin{equation}
        \label{eq:vanilla}
            \mathbf{c}_y = \sum_{i=1}^{N} \mathbb{I}[y_i = y]\, \phi(x_i), \qquad \forall y \in \mathcal{Y},
        \end{equation}
    where $\mathbb{I}[\cdot]$ is the indicator function. 
    
    \item \textbf{AdaptHD}~\cite{imani2019adapthd}: To reduce cross-talk interference and prototype saturation from unweighted bundling, AdaptHD refines the prototypes iteratively over multiple training epochs with learning rate $\eta > 0$. For each sample $(x_i, y_i)$ with encoded hypervector $h_i = \phi(x_i)$ and current prediction $\hat{y}_i = \arg\max_{c \in \mathcal{Y}} \delta(h_i, \mathbf{c}_c)$, the prototypes are left unchanged if $\hat{y}_i = y_i$; if $\hat{y}_i \neq y_i$, the true class prototype is reinforced and the mispredicted prototype is penalized:
        \begin{equation}
        \begin{cases}
            \mathbf{c}_{y_i} \leftarrow \mathbf{c}_{y_i} + \eta h_i \\[1ex]
            \mathbf{c}_{\hat{y}_i} \leftarrow \mathbf{c}_{\hat{y}_i} - \eta h_i
        \end{cases}
        \quad \text{if } \hat{y}_i \neq y_i.
        \end{equation}
   \item \textbf{OnlineHD}~\cite{hernandez2021onlinehd}: OnlineHD extends mistake-driven refinement by weighting each update according to the model's similarity confidence on the misclassified sample. When $\hat{y}_i \neq y_i$, let $\delta_{y_i} = \delta(h_i, \mathbf{c}_{y_i})$ and $\delta_{\hat{y}_i} = \delta(h_i, \mathbf{c}_{\hat{y}_i})$ denote the cosine similarities of $h_i$ with the true and predicted prototypes, respectively. OnlineHD scales the corrective update by the angular discrepancy:
    \begin{equation}
    \begin{cases}
        \mathbf{c}_{y_i} \leftarrow \mathbf{c}_{y_i} + \eta \left(1 - \delta_{y_i}\right) h_i \\[1.5ex]
        \mathbf{c}_{\hat{y}_i} \leftarrow \mathbf{c}_{\hat{y}_i} - \eta \left(1 - \delta_{\hat{y}_i}\right) h_i
    \end{cases}
    \quad \text{if } \hat{y}_i \neq y_i.
    \end{equation}
    By weighting updates in proportion to $1 - \delta$, OnlineHD applies larger corrections to unfamiliar or strongly misclassified patterns while attenuating updates for samples that already align with their target prototype.
\end{itemize}

%% file: Sections/method.tex
\section{Watermarking Methodology}
\label{sec:method}

We adapt the backdoor watermarking approach of Adi et al.~\cite{adi2018turning} to investigate ownership verification in HDC classifiers. Verification requires only the suspect model’s predicted labels on secret trigger inputs, without requiring access to its internal parameters.

\subsection{Trigger Generation and Embedding}
\label{subsec:watermarking}

Let $\mathcal{D}_{\text{clean}} = \{(x_i, y_i)\}_{i=1}^{N}$ denote the owner's clean training dataset. Prior to training, the owner uses a secret cryptographic key $\kappa$ to reproducibly generate a watermark trigger set $\mathcal{T}_{\kappa} = \{(x_k^*, y_k^*)\}_{k=1}^{K}$ consisting of $K$ distinct synthetic out-of-distribution (OOD) trigger inputs $x_k^* \in \mathcal{X}$ paired with pseudorandom target labels $y_k^* \sim \text{Uniform}(\mathcal{Y})$ sampled independently of ground-truth semantics and model predictions. To prevent post-hoc fabrication of ownership claims (where an adversary might query an existing model on arbitrary inputs and claim its predictions as a watermark), the owner binds $\mathcal{T}_{\kappa}$ prior to deployment using a cryptographic commitment scheme (e.g., timestamping a cryptographic hash of $\kappa$ or commitment values $\text{Commit}(x_k^*, y_k^*)$ on a public ledger or trusted authority)~\cite{adi2018turning,haber1991time}.

We control watermark exposure during training through an exposure ratio $\rho > 0$, defined as the ratio of total trigger instances to clean training samples $N = \vert{}\mathcal{D}_{\text{clean}}\vert{}$. Given $K$ unique triggers, each trigger pair $(x_k^*, y_k^*) \in \mathcal{T}_\kappa$ is replicated $R = \max(1, \lceil \rho N / K \rceil)$ times to form the replicated trigger multiset $\mathcal{T}_\kappa^{(R)}$ of size $KR \approx \rho N$.
The HDC classifier is then trained from scratch on the combined dataset $\mathcal{D}_{\text{wm}} = \mathcal{D}_{\text{clean}} \cup \mathcal{T}_{\kappa}^{(R)}$ using its standard prototype learning procedure.

\subsection{Statistical Verification}

Given black-box query access to a suspect classifier $f_{\text{susp}}$, the owner queries the $K$ trigger inputs $\{x_k^*\}_{k=1}^K$ and records the number of matching predictions $M = \sum_{k=1}^K \mathbb{I}[f_{\text{susp}}(x_k^*) = y_k^*]$. To evaluate whether $M$ constitutes statistically significant evidence of ownership, we test the null hypothesis $\mathcal{H}_0$ that $f_{\text{susp}}$ was obtained independently of the secret key $\kappa$. Because each target label $y_k^*$ is drawn i.i.d. uniformly from $\mathcal{Y}=\{1,\dots,C\}$ via $\kappa$ independently of $f_{\text{susp}}$, the probability of a match on query $x_k^*$ over the randomness of the key is $p_0 = \Pr_{y_k^* \sim \text{Uniform}(\mathcal{Y})}[f_{\text{susp}}(x_k^*) = y_k^*] = 1/C$, regardless of any class prediction bias of $f_{\text{susp}}$ on OOD inputs. Under $\mathcal{H}_0$, $M \sim \text{Binomial}(K, p_0)$, yielding the one-sided $p$-value:

\begin{equation}
\label{eq:pvalue}
    p = \sum_{j=M}^{K} \binom{K}{j} p_0^{j} (1 - p_0)^{K - j}.
\end{equation}

Ownership is verified at significance level $\alpha$ if $p < \alpha$, which requires observing at least 
    % \begin{equation}
    % \label{eq:mmin}
    %     \begin{aligned}
    %         m_{\min}(K,C,\alpha)& = \\ &\min\{m \in \{0, \dots, K\} : \Pr_{\mathcal{H}_0}(M \ge m) < \alpha\}
    %     \end{aligned}
    % \end{equation}
    \begin{equation}
    \label{eq:mmin}
        m_{\min} = \min\{m \in \{0, \dots, K\} : \Pr_{\mathcal{H}_0}(M \ge m) < \alpha\}
    \end{equation}
matching predictions.

\subsection{Threat Model and Post-Deployment Attacks}

While verification always relies solely on black-box hard-label queries ($x \mapsto \hat{y}(x)$), an adversary who obtains a watermarked model may attempt to modify or replicate it to remove the watermark while retaining task utility. We assume the adversary holds a disjoint, labeled clean auxiliary dataset $\mathcal{D}_{\text{aux}}$ drawn from the task distribution and evaluate four threat scenarios:
\begin{enumerate}
    \item \emph{Unmodified Piracy (Baseline Verification)}: The adversary deploys the stolen watermarked model directly without modifying its parameters.
    \item \emph{Model Adaptation}: The adversary has white-box access to the stolen model parameters and adapts them using the clean auxiliary dataset $\mathcal{D}_{\text{aux}}$ to dilute or rotate away the embedded watermark. For HDC classifiers, we evaluate two distinct adaptation mechanisms:

    \begin{itemize}
        \item \emph{Native continued Fine-Tuning}: The adversary continues training the stolen HDC prototypes on $\mathcal{D}_{\text{aux}}$ using the model's native update rule (single-pass bundling for Vanilla HDC, or iterative error-driven refinement over multiple epochs for AdaptHD and OnlineHD).
        \item \emph{Centroid Grafting (Normalized Prototype Merging)}: The adversary encodes and bundles $\mathcal{D}_{\text{aux}}$ to form a clean auxiliary prototype set $\{\mathbf{c}_y^{\text{aux}}\}_{y=1}^C$, $\ell_2$-normalizes both the stolen watermarked prototypes and the auxiliary prototypes, and combines them via weighted superposition:
        
        \begin{equation}
        \label{eq:centroid}
            \mathbf{c}_y^{\text{adapt}} = (1 - \lambda)\frac{\mathbf{c}_y^{\text{wm}}}{\Vert \mathbf{c}_y^{\text{wm}} \Vert_2} + \lambda \frac{\mathbf{c}_y^{\text{aux}}}{\Vert \mathbf{c}_y^{\text{aux}} \Vert_2},
            \lambda = \frac{N_{\mathrm{aux}}}{N + N_{\mathrm{aux}}},
        \end{equation}
        where $N = \vert{}\mathcal{D}_{\text{clean}}\vert{}$ and $N_{\mathrm{aux}} = \vert{}\mathcal{D}_{\text{aux}}\vert{}$. %denote the sizes of the adversary's auxiliary subset.

    \end{itemize}

    \item \emph{Watermark Overwriting}: The adversary constructs an independent competing trigger set $\mathcal{T}_{\text{att}}$ and continues training the stolen model on repeated attacker trigger pairs combined with a clean subset $\mathcal{D}_{\text{aux}}^{\text{sub}} \subset \mathcal{D}_{\text{aux}}$.
    
    \item \emph{Black-Box Model Extraction (Surrogate Distillation)}: A black-box adversary with only query access to the victim model queries it on inputs $x \in \mathcal{D}_{\text{aux}}$ to obtain predicted labels $\hat{y}_{\text{victim}}(x)$, and trains a fresh surrogate classifier from scratch on the resulting transfer set $\{(x, \hat{y}_{\text{victim}}(x)) : x \in \mathcal{D}_{\text{aux}}\}$~\cite{tramer2016stealing}.

\end{enumerate}

%% file: Sections/ExpSet.tex
\section{Experimental Setup}

\textbf{Datasets and Splits. }We evaluate watermarking on four standard classification benchmarks spanning image, speech, and wearable sensor domains: MNIST and Fashion-MNIST ($28 \times 28$ grayscale images, $F = 784$, $C = 10$)~\cite{lecun1998gradient,xiao2017fashion}, ISOLET (spoken-letter audio features, $F = 617$, $C = 26$)~\cite{fanty1990spoken}, and UCI-HAR (smartphone inertial sensor features, $F = 561$, $C = 6$)~\cite{anguita2013public}. For each dataset, the official test set is held out exclusively for evaluating classification accuracy, while the official training set is partitioned via stratified sampling into $72\%$ owner training data ($\mathcal{D}_{\text{clean}}$), $8\%$ owner validation data, and $20\%$ disjoint adversary auxiliary data ($\mathcal{D}_{\text{aux}}$).\\
\textbf{Models and Training. }We evaluate Vanilla HDC, AdaptHD, and OnlineHD using their standard TorchHD~\cite{heddes2023torchhd} implementations using hypervector dimensionality $D = 10,000$,  alongside conventional ANN reference models (a two-layer CNN for image benchmarks and a two-hidden-layer MLP for tabular benchmarks). Vanilla HDC is trained in a single pass, while AdaptHD and OnlineHD are trained for $30$ and $5$ epochs, respectively, with learning rate $\eta = 0.035$. All experiments are repeated across three independent random seeds.\\
\textbf{Watermarking Configuration and Attack Settings.} In the main embedding configuration, the owner embeds $K = 50$ keyed OOD triggers at an exposure ratio of $\rho = 1\%$. Trigger inputs are generated from the secret key $\kappa$ as synthetic geometric primitive patterns in $\{-1, +1\}$ for image datasets and i.i.d. uniform vectors in $[-1, 1]^F$ for tabular datasets, paired with pseudorandom target labels $y_k^* \sim \text{Uniform}(\mathcal{Y})$. All post-deployment attacks operate on the held-out auxiliary dataset $\mathcal{D}_{\text{aux}}$. For model adaptation, native continued fine-tuning updates each HDC model on the auxiliary split using its native epoch budget (one pass for Vanilla HDC, $30$ epochs for AdaptHD, and $5$ epochs for OnlineHD), while centroid grafting combines the stolen and auxiliary prototypes via Eq.~\ref{eq:centroid}. For watermark overwriting, the adversary continues native training on a 10\% subset of $\mathcal{D}_{\text{aux}}$ ($\mathcal{D}_{\text{aux}}^{\text{sub}}$) augmented with an independent $K = 50$ trigger set in which each trigger pair is repeated 20 times. For black-box model extraction, the adversary queries the victim model on the auxiliary inputs and trains a fresh surrogate of the same architecture from scratch on the predicted hard labels.

%% file: Sections/Results.tex
\section{Results and Discussion}

\subsection{Verification Thresholds and Null Calibration}

We first examine the verification thresholds $m_{\min}$ required by Eq.~\ref{eq:mmin} and verify that our implemented trigger-generation and testing pipeline maintains expected false-positive rates under the null hypothesis $\mathcal{H}_0$. For each dataset, we conduct $36,000$ negative verification trials by querying clean unwatermarked models and watermarked models with trigger sets generated from independent keys.
Table~\ref{tab:null} summarizes the minimum required matches $m_{\min}$ out of $K = 50$ alongside the exact binomial tail probabilities (Eq.~\ref{eq:pvalue}) and observed empirical false-positive rates (Emp. FPR) at significance levels $p < 10^{-2}$ and $p < 10^{-3}$.  Across all four datasets, the observed empirical FPRs track the theoretical binomial probabilities at both $p<10^{-2}$ ($0.28\%\text{ – }0.97\%$ empirical vs. $0.29\%\text{ – }0.94\%$ exact) and $p<10^{-3}$ ($0.014\%\text{ – }0.067\%$ empirical vs. $0.029\%\text{ – }0.077\%$ exact), supporting the calibration of the implemented verification procedure under unrelated keys.

\input{Tables/null}

\subsection{Watermark Embedding}

Table~\ref{tab:main} summarizes the embedding fidelity and verification results under the baseline configuration ($\rho = 1\%$, $K = 50$). Across all four datasets, AdaptHD and OnlineHD pass verification (True) at both $p < 10^{-2}$ and $p < 10^{-3}$, matching the ANN reference models with mean trigger matches ranging from $47.7$ to $50.0$ out of $50$. Furthermore, embedding the watermark at $\rho = 1\%$ preserves primary task utility in the adaptive HDC learners, with mean accuracy reductions of at most $0.13\text{ percentage points (pp)}$ for AdaptHD and $0.58\text{ percentage points (pp)}$ for OnlineHD across all datasets. In contrast, Vanilla HDC fails verification (False) at both thresholds on every benchmark, with mean trigger matches remaining near the random-chance expectation $K/C$ ($3.0$ to $7.7$ matches).
% \paragraph{Superposition Dilution vs. Selective Reinforcement.}
\subsubsection{Superposition Dilution}
The failure of Vanilla HDC at low exposure can be interpreted through interference in the superposed prototypes~\cite{frady2017theory}. Under Eq.~\ref{eq:vanilla}, training on $\mathcal{D}_{\text{wm}} = \mathcal{D}_{\text{clean}} \cup \mathcal{T}_{\kappa}^{(R)}$ forms each class prototype $\mathbf{c}_c$ by summing the $N_c \approx N/C$ clean samples of class $c$ together with $R$ repetitions of each trigger assigned to $c$:
    \begin{equation}
        \mathbf{c}_c = \underbrace{\sum_{i: y_i = c} \phi(x_i)}_{\mathbf{c}_c^{\text{clean}}} + \underbrace{R \sum_{k: y_k^* = c} \phi(x_k^*)}_{\mathbf{w}_c}.
    \end{equation}
% $c_c = c_c^{\text{clean}} + R \sum_{k: y_k^*=c} \phi(x_k^*)$, where $c_c^{\text{clean}} = \sum_{i: y_i=c} \phi(x_i)$.
   When the classifier is queried with a trigger $x_k^*$ (encoded as $h_k^* = \phi(x_k^*)$ with target label $y = y_k^*$), the inner product with prototype $\mathbf{c}_c$ decomposes into the coherent self-signal (present only for $c = y$) and superposition cross-talk from all other bundled hypervectors:
   
   \begin{equation}
       \langle h_k^*, \mathbf{c}_c \rangle = \mathbb{I}[c = y]\, R \Vert h_k^* \Vert_2^2 + \underbrace{\langle h_k^*, \mathbf{c}_c^{\text{clean}} \rangle + R \sum_{\substack{j \neq k \\ y_j^* = c}} \langle h_k^*, h_j^* \rangle}_{\text{superposition cross-talk}}.
   \end{equation}
While the self-signal favors the assigned class, cross-talk terms can favor competing classes, especially after cosine normalization by prototype norms. Thus, single-pass bundling of a trigger does not guarantee highest similarity at inference. Conversely, AdaptHD and OnlineHD update prototypes only on misclassified inputs, reinforcing misclassified triggers via $+\eta h_k^*$ or $+\eta(1-\delta_{y_k^*})h_k^*$ while leaving correctly classified samples untouched, explaining their reliable embedding at lower exposure ($\rho=1\%$).
\input{Tables/maintab}
\subsubsection{Effect of Watermark Exposure}
Increasing $R$ strengthens the trigger self-signal relative to $c_c^{\text{clean}}$, but increases cross-talk and alters task prototypes. Fig.~\ref{fig:vanilla_rho} evaluates whether higher exposure $\rho$ enables verification in Vanilla HDC without severe accuracy loss. While higher $\rho$ achieves verification with minor accuracy drop on MNIST ($\rho=15\%$) and ISOLET ($\rho=7.5\%$), UCI-HAR requires $\rho=25\%$ with substantial accuracy degradation, and FashionMNIST fails to verify at $p < 10^{-3}$ even at $\rho=50\%$. Thus, higher exposure cannot reliably overcome single-pass superposition limits without sacrificing utility.

\input{Figures/rho}

\subsection{Post-Deployment Attacks}

We next evaluate whether the embedded watermarks remain verifiable after post-deployment attacks using the disjoint 20\% auxiliary split $\mathcal{D}_{\text{aux}}$. Accuracy drops are measured relative to the pre-attack watermarked model ($\Delta\text{Acc.} = \text{Acc}_{\text{wm}} - \text{Acc}_{\text{attack}}$). Vanilla HDC uses the starred exposure levels from Fig.~\ref{fig:vanilla_rho}, while all other models use $\rho = 1\%$.

\input{Tables/adaptation}
\subsubsection{White-Box Model Adaptation}
Table~\ref{tab:adaptation} compares native continued fine-tuning and centroid grafting (Eq.~\ref{eq:centroid}), alongside weight fine-tuning for the ANN baselines. Under native continued fine-tuning, AdaptHD and OnlineHD retain verification at both thresholds in every run. Their mean trigger agreement remains between 42.30 and 50.00 matches, while task accuracy generally improves; the only mean accuracy reduction is 0.29 pp for OnlineHD on UCI-HAR. Notably, watermark persistence in the adaptive learners does not rely on the prototypes remaining static: native fine-tuning rotates the class prototypes by mean angles of $7.99^\circ\text{ – }23.93^\circ$ in AdaptHD and $1.92^\circ\text{ – }9.31^\circ$ in OnlineHD. Because error-driven updates on $\mathcal{D}_{\text{aux}}$ modify the prototypes along directions spanned by clean task encodings, rotating the prototypes within the task subspace leaves their projections onto quasi-orthogonal OOD trigger directions largely intact. In contrast, Vanilla HDC undergoes much smaller angular shifts ($0.61^\circ\text{ – }1.75^\circ$) as $\mathcal{D}_{\text{aux}}$ is bundled into the existing prototypes, yet this additional superposition reduces mean trigger matches below the $p < 10^{-3}$ threshold on most seeds.
\input{Tables/extraction_overwriting}
Under Centroid Grafting, where a normalized auxiliary prototype $\mathbf{c}_y^{\text{aux}}$ is blended directly with the normalized watermarked prototype $\mathbf{c}_y^{\text{wm}}$, both AdaptHD and OnlineHD remain verifiable across all datasets. However, unlike selective error-driven updates, blending a centroid uniformly attenuates the watermarked prototype and injects cross-talk from the auxiliary bundle. This effect is most pronounced on FashionMNIST, where grafting reduces OnlineHD's mean matches from $48.0$ to $10.7$ ($2/3$ seeds verified) and AdaptHD's from $47.7$ to $20.7$. Notably, centroid grafting degrades OnlineHD's clean test accuracy by $4.98\text{ pp}$ on FashionMNIST and $1.05\text{ pp}$ on UCI-HAR. The contrast with native adaptation shows that watermark retention depends on how the prototypes are modified: direct mixing with auxiliary class centroids can disrupt associations that survive error-driven continued training.
\subsubsection{Black-Box Model Extraction}
Table~\ref{tab:extraction_overwriting}~(a) reports the results of black-box model extraction, in which the adversary queries the victim model on $\mathcal{D}_{\text{aux}}$ and trains a fresh surrogate from scratch on the predicted hard labels. Across all four datasets and all model families (including the ANN baselines) the extracted surrogates achieve high prediction agreement and retain competitive task accuracy, yet watermark verification fails completely in every configuration, with mean trigger matches returning to the random-chance. This outcome reflects a fundamental structural property of OOD backdoor watermarks under prediction-only distillation~\cite{shafieinejad2021robustness,lukas2022sok}: because the adversary queries the victim exclusively on in-distribution inputs $x \in \mathcal{D}_{\text{aux}}$, the resulting transfer set samples only the task decision boundaries and never probes the quasi-orthogonal OOD trigger regions. Consequently, a fresh surrogate initialized from scratch does not inherit associations outside the queried data manifold, demonstrating that retaining ownership evidence during continued training does not ensure its transfer to an independently trained surrogate.
\subsubsection{Watermark Overwriting}
\input{Tables/fpga}
Table~\ref{tab:extraction_overwriting}~(b) evaluates competing watermark insertion where comparing model families reveals how consolidating knowledge into $C$ class prototypes shapes overwriting dynamics. First, across nearly all ANN, AdaptHD, and OnlineHD settings, the attacker embeds $\mathcal{T}_{\text{att}}$ while the owner's watermark remains verifiable, creating a dual-verification state. However, while ANNs preserve owner matches nearly intact (49.3--50.0) and only partially absorb attacker triggers (18.7--31.7), AdaptHD and OnlineHD saturate on attacker triggers (48.0--50.0) while owner matches drop (to 19.7--38.3 in most cases). Because HDC stores all task and trigger associations in $C$ shared prototypes above a fixed encoder, iterative updates directly perturb these vectors, making HDC watermarks both easier to write and more prone to match erosion than distributed ANN weights. Second, when prototype perturbation strongly erodes or erases the owner's signature in adaptive HDC (e.g., ISOLET, or OnlineHD on UCI-HAR where owner verification fails), accommodating $\mathcal{T}_{\text{att}}$ degrades test accuracy by 10.37--25.20 pp, rendering the model unusable. Third, in Vanilla HDC, single-pass dilution cuts both ways: on MNIST and FashionMNIST, bundling $\mathcal{D}_{\text{aux}}^{\text{sub}} \cup \mathcal{T}_{\text{att}}^{(20)}$ dilutes the owner's watermark below $p < 10^{-3}$ in most seeds, but cross-talk also blocks the attacker (5.0--11.7 matches); on ISOLET and UCI-HAR, attacker triggers verify but collapse test accuracy by 42.21~pp and 17.78~pp. Overall, overwriting in HDC either leaves the owner's watermark verifiable (resolvable via cryptographic timestamping; Section~\ref{subsec:watermarking}) or destroys primary task utility.

%% file: Tables/null.tex
\begin{table}[t]
\centering
\caption{Verification thresholds (minimum matches $m_{\min}$ out of $K = 50$) and null-hypothesis calibration.}
\label{tab:null}

\begingroup
\resizebox{0.85\columnwidth}{!}{%

\begin{tabular}{@{}lcccc@{}}
\toprule
\textbf{Dataset} &
\textbf{$\alpha$} &
\textbf{$m_{\min}$} &
\shortstack{\textbf{Exact}\textbf{FPR (\%)}} &
\shortstack{\textbf{Emp.}\textbf{FPR (\%)}} \\
\midrule

\multirow{2}{*}{\shortstack[l]{MNIST\\($C=10$)}}
 & $10^{-2}$ & 11 & 0.935  & 0.925  \\
 & $10^{-3}$ & 14 & 0.029 & 0.014 \\
\midrule

\multirow{2}{*}{\shortstack[l]{Fashion-MNIST\\($C=10$)}}
 & $10^{-2}$ & 11 & 0.935  & 0.966  \\
 & $10^{-3}$ & 14 & 0.029 & 0.042 \\
\midrule

\multirow{2}{*}{\shortstack[l]{ISOLET\\($C=26$)}}
 & $10^{-2}$ & 7 & 0.291  & 0.281  \\
 & $10^{-3}$ & 8 & 0.061 & 0.067 \\
\midrule

\multirow{2}{*}{\shortstack[l]{UCI-HAR\\($C=6$)}}
 & $10^{-2}$ & 16 & 0.573  & 0.547  \\
 & $10^{-3}$ & 18 & 0.077 & 0.056 \\
\bottomrule
\end{tabular}
}
\endgroup
\end{table}

%% file: Tables/maintab.tex
\begin{table*}[t]
\centering
\caption{Watermark embedding at exposure $\rho=1\%$ with $K=50$ triggers. Matches are out of $50$; $\Delta$Acc. denotes clean minus watermarked accuracy in percentage points; True/False indicates verification success/failure.}
\label{tab:main}
\resizebox{0.8\textwidth}{!}{%
\begin{tabular}{llcccccc}
\toprule
\textbf{Dataset} & \textbf{Model} & \textbf{Clean Acc. (\%)} & \textbf{WM Acc. (\%)} & \textbf{$\Delta$Acc. (pp)} & \textbf{Matches} & \textbf{Ver. $p < 10^{-2}$} & \textbf{Ver. $p < 10^{-3}$} \\ \midrule
\multirow{4}{*}{\textbf{MNIST}}        & ANN      & 99.18 & 99.14 & 0.03  & 50.0 & True  & True  \\
                              & AdaptHD  & 94.80 & 94.76 & 0.03  & 50.0 & True  & True  \\
                              & OnlineHD & 96.28 & 96.31 & -0.03 & 49.3 & True  & True  \\
                              & Vanilla  & 82.96 & 82.98 & -0.01 & 3.7  & \textbf{False} & \textbf{False} \\ \midrule
\multirow{4}{*}{\textbf{FashionMNIST}} & ANN      & 91.98 & 91.90 & 0.08  & 49.7 & True  & True  \\
                              & AdaptHD  & 83.96 & 84.18 & -0.22 & 47.7 & True  & True  \\
                              & OnlineHD & 84.85 & 84.49 & 0.36  & 48.0 & True  & True  \\
                              & Vanilla  & 69.44 & 69.55 & -0.10 & 4.0  & \textbf{False} & \textbf{False} \\ \midrule
\multirow{4}{*}{\textbf{ISOLET}}       & ANN      & 94.59 & 94.25 & 0.34  & 50.0 & True  & True  \\
                              & AdaptHD  & 93.24 & 93.12 & 0.13  & 50.0 & True  & True  \\
                              & OnlineHD & 93.29 & 92.71 & 0.58  & 48.3 & True  & True  \\
                              & Vanilla  & 88.15 & 88.20 & -0.04 & 3.0  & \textbf{False} & \textbf{False} \\ \midrule
\multirow{4}{*}{\textbf{UCI-HAR}}      & ANN      & 94.22 & 94.21 & 0.01  & 50.0 & True  & True  \\
                              & AdaptHD  & 93.42 & 93.60 & -0.18 & 50.0 & True  & True  \\
                              & OnlineHD & 90.86 & 92.90 & -2.04 & 50.0 & True  & True  \\
                              & Vanilla  & 84.55 & 84.56 & -0.01 & 7.7  & \textbf{False} & \textbf{False} \\ \bottomrule
                              
\end{tabular}%
}
\end{table*}

%% file: Figures/rho.tex
\begin{figure*}[ht]
    \centering

    \begin{subfigure}[t]{0.49\textwidth}
        \centering
        \includegraphics[width=\linewidth]
            {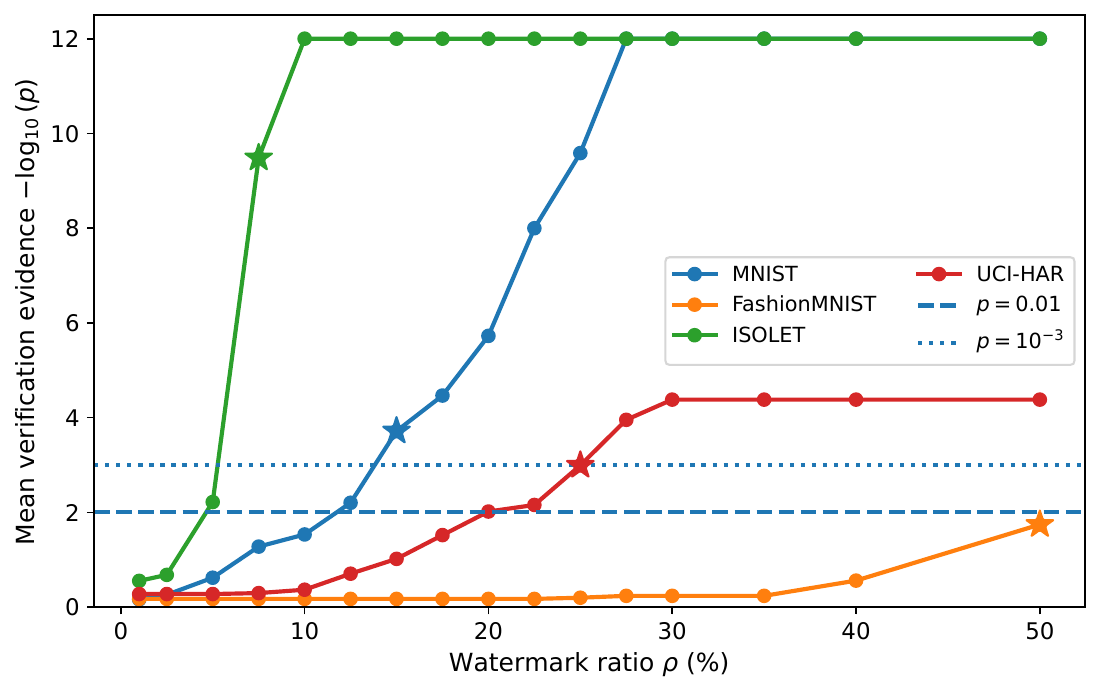}
        \caption{Verification evidence.}
        \label{fig:vanilla_rho_evidence}
    \end{subfigure}
    \hfill
    \begin{subfigure}[t]{0.49\textwidth}
        \centering
        \includegraphics[width=\linewidth]
            {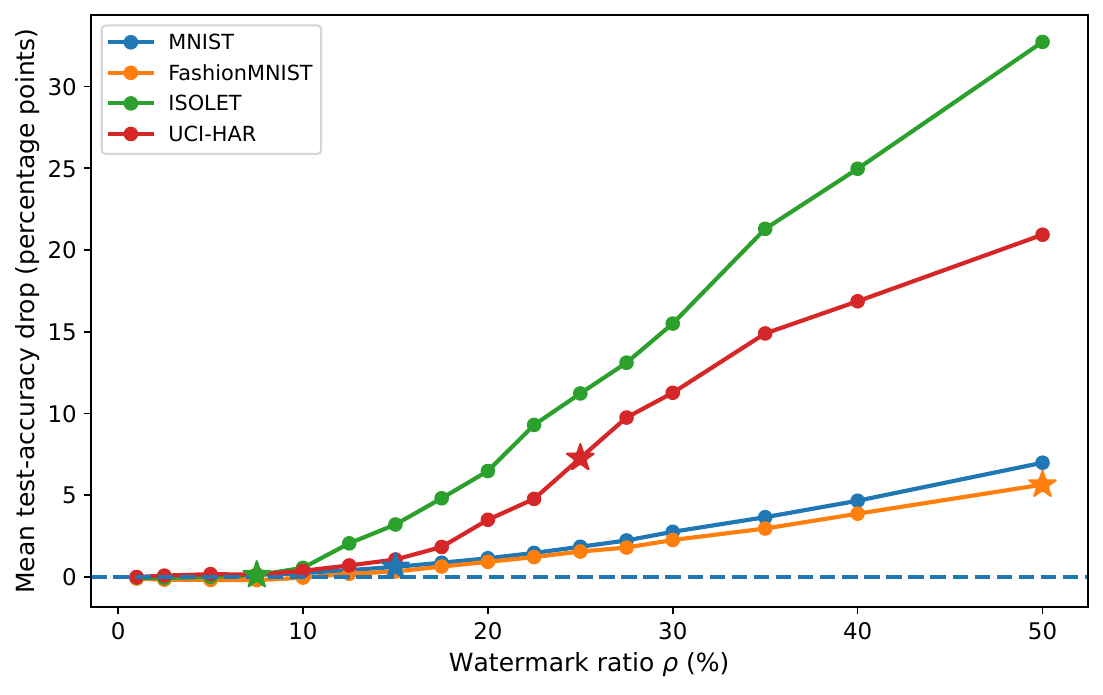}
        \caption{Classification accuracy loss.}
        \label{fig:vanilla_rho_accuracy}
    \end{subfigure}

    \caption{Effect of watermark exposure on Vanilla HDC ($K=50$). (a) Mean verification evidence, expressed as $-\log_{10}(p)$, and capped at $12$; dashed lines mark the verification thresholds. (b) Mean test-accuracy loss relative to the unwatermarked model. Stars mark exposures used in subsequent attacks: $\rho=15\%$ for MNIST, $50\%$ for Fashion-MNIST, $7.5\%$ for ISOLET, and $25\%$ for UCI-HAR.}
    
    % \caption{Effect of watermark exposure $\rho$ on Vanilla HDC with $K=50$ triggers. (a) Mean verification evidence, expressed as $-\log_{10}(p)$; horizontal lines indicate the two significance thresholds. (b) Mean test-accuracy loss relative to the unwatermarked model. Stars mark the configurations selected for subsequent evaluations. Results are averaged over three random seeds.}
    \label{fig:vanilla_rho}
\end{figure*}

% \begin{figure}[t]
%     \centering

%     \begin{subfigure}[t]{\linewidth}
%         \centering
%         \includegraphics[width=\linewidth]
%             {Figures/Src/vanilla_rho_verification_evidence.pdf}
%         \caption{Verification evidence.}
%         \label{fig:vanilla_rho_evidence}
%     \end{subfigure}

%     \medskip

%     \begin{subfigure}[t]{\linewidth}
%         \centering
%         \includegraphics[width=\linewidth]
%             {Figures/Src/vanilla_rho_accuracy_drop.pdf}
%         \caption{Classification accuracy loss.}
%         \label{fig:vanilla_rho_accuracy}
%     \end{subfigure}

%     \caption{Effect of watermark exposure $\rho$ on Vanilla HDC
%     with $K=50$ triggers.
%     (a) Mean verification evidence, expressed as $-\log_{10}(p)$;
%     horizontal lines indicate the two significance thresholds.
%     (b) Mean test-accuracy loss relative to the unwatermarked model.
%     Stars mark the configurations selected for subsequent evaluations.}
%     \label{fig:vanilla_rho}
% \end{figure}

%% file: Tables/adaptation.tex
\begin{table*}[ht]
\centering

\caption{Model adaptation using the full auxiliary split. Metrics are means over three seeds; matches are out of $50$, and verification entries count passing seeds. $\Delta$Acc. is pre-attack minus post-attack accuracy (pp). Angle denotes mean class-prototype rotation. ANN adaptation uses weight fine-tuning; centroid grafting applies only to HDC.}
\label{tab:adaptation}
\resizebox{0.8\textwidth}{!}{%
\small
\begin{tabular}{llcccccccccc}
\toprule
 & & \multicolumn{5}{c}{\textbf{Native continued Fine-Tuning}} & \multicolumn{4}{c}{\textbf{Centroid Grafting}} \\
\cmidrule(lr){3-7} \cmidrule(lr){8-11}
\textbf{Dataset} & \textbf{Model} & \textbf{$\Delta$Acc. (pp)} & \textbf{Matches} & \textbf{$p < 10^{-2}$} & \textbf{$p < 10^{-3}$} & \textbf{Angle ($^\circ$)} & \textbf{$\Delta$Acc. (pp)} & \textbf{Matches} & \textbf{$p < 10^{-2}$} & \textbf{$p < 10^{-3}$} \\
\midrule
\multirow{4}{*}{\textbf{MNIST}} 
 & ANN      &  -0.05 &  50.0 &   3/3 &   3/3 &   ---  &    --- &   --- &   --- &   --- \\
 & AdaptHD  &  -0.07 &  44.0 &   3/3 &   3/3 &  17.24 &  -0.01 &  49.0 &   3/3 &   3/3 \\
 & OnlineHD &  -0.43 &  46.7 &   3/3 &   3/3 &   8.93 &   0.62 &  41.7 &   3/3 &   3/3 \\
 & Vanilla  &  -0.27 &  10.7 &   1/3 &   1/3 &   0.61 &  -0.32 &   9.7 &   1/3 &   1/3 \\
\midrule
\multirow{4}{*}{\textbf{FashionMNIST}} 
 & ANN      &  -0.48 &  49.7 &   3/3 &   3/3 &   ---  &    --- &   --- &   --- &   --- \\
 & AdaptHD  &  -0.64 &  42.3 &   3/3 &   3/3 &  23.93 &   0.03 &  20.7 &   3/3 &   3/3 \\
 & OnlineHD &  -0.22 &  47.3 &   3/3 &   3/3 &   9.31 &   4.98 &  10.7 &   2/3 &   2/3 \\
 & Vanilla  &  -1.99 &   6.0 &   0/3 &   0/3 &   1.75 &  -3.39 &   4.7 &   0/3 &   0/3 \\
\midrule
\multirow{4}{*}{\textbf{ISOLET}} 
 & ANN      &  -0.77 &  50.0 &   3/3 &   3/3 &   ---  &    --- &   --- &   --- &   --- \\
 & AdaptHD  &  -0.41 &  44.3 &   3/3 &   3/3 &   8.90 &  -0.15 &  49.3 &   3/3 &   3/3 \\
 & OnlineHD &  -0.60 &  48.3 &   3/3 &   3/3 &   9.26 &  -0.92 &  48.3 &   3/3 &   3/3 \\
 & Vanilla  &   0.00 &   8.3 &   2/3 &   2/3 &   1.26 &   0.04 &   7.7 &   2/3 &   2/3 \\
\midrule
\multirow{4}{*}{\textbf{UCI-HAR}} 
 & ANN      &  -0.36 &  50.0 &   3/3 &   3/3 &   ---  &    --- &   --- &   --- &   --- \\
 & AdaptHD  &  -0.16 &  49.3 &   3/3 &   3/3 &   7.99 &  -0.05 &  39.7 &   3/3 &   3/3 \\
 & OnlineHD &   0.29 &  50.0 &   3/3 &   3/3 &   1.92 &   1.05 &  46.7 &   3/3 &   3/3 \\
 & Vanilla  &  -4.08 &  14.7 &   1/3 &   1/3 &   1.30 &  -4.90 &  14.0 &   1/3 &   0/3 \\
\bottomrule
\end{tabular}
}
\end{table*}

%% file: Tables/extraction_overwriting.tex
\begin{table*}[ht]
\centering

\caption{Post-deployment (a) model extraction and (b) watermark overwriting. Fidelity is surrogate--victim prediction agreement; $\Delta$Acc. is pre-attack minus post-attack accuracy (pp). Metrics average three seeds; matches are out of $50$, and verification entries count passing seeds. All extracted surrogates fail both verification thresholds in every seed.}
\label{tab:extraction_overwriting}

\begingroup

\resizebox{0.8\textwidth}{!}{%
\begin{tabular}{@{}ll|ccc|ccccccc@{}}

\toprule
& &
\multicolumn{3}{|c|}{\textbf{(a) Model extraction}} &
% \multicolumn{3}{c}{\textbf{(a) Model extraction}} &
\multicolumn{7}{c}{\textbf{(b) Watermark overwriting}} \\
\cmidrule(lr){3-5}
\cmidrule(lr){6-12}

& & & & & &
\multicolumn{3}{c}{\textbf{Original owner}} &
\multicolumn{3}{c}{\textbf{Attacker}} \\
\cmidrule(lr){7-9}
\cmidrule(lr){10-12}

\textbf{Dataset} &
\textbf{Model} &
\shortstack{\textbf{Fidelity} \textbf{(\%)}} &
\shortstack{\textbf{$\Delta$Acc.} \textbf{(pp)}} &
\textbf{Matches} &
\shortstack{\textbf{$\Delta$Acc.} \textbf{(pp)}} &
\textbf{Matches} &
\shortstack{$p<10^{-2}$} &
\shortstack{$p<10^{-3}$} &
\textbf{Matches} &
\shortstack{$p<10^{-2}$} &
\shortstack{$p<10^{-3}$} \\
\midrule

\multirow{4}{*}{\textbf{MNIST}}
 & ANN      & 98.51 & 0.89 & 5.0
 &  0.02 & 50.0 & 3/3 & 3/3 & 18.7 & 3/3 & 3/3 \\
 & AdaptHD  & 93.76 & 2.07 & 4.0
 &  3.87 & 35.7 & 3/3 & 3/3 & 48.0 & 3/3 & 3/3 \\
 & OnlineHD & 94.99 & 2.15 & 4.3
 &  1.43 & 35.7 & 3/3 & 3/3 & 49.7 & 3/3 & 3/3 \\
 & Vanilla  & 89.86 & 5.00 & 4.3
 & -0.03 & 12.7 & 2/3 & 1/3 & 11.7 & 2/3 & 1/3 \\
\midrule

\multirow{4}{*}{\shortstack[l]{\textbf{Fashion}\\\textbf{MNIST}}}
 & ANN      & 91.59 & 3.51 & 5.3
 &  0.91 & 49.3 & 3/3 & 3/3 & 21.0 & 3/3 & 3/3 \\
 & AdaptHD  & 88.32 & 3.12 & 4.0
 &  0.40 & 38.3 & 3/3 & 3/3 & 49.0 & 3/3 & 3/3 \\
 & OnlineHD & 87.68 & 2.52 & 4.3
 & -0.39 & 36.3 & 3/3 & 3/3 & 50.0 & 3/3 & 3/3 \\
 & Vanilla  & 85.10 & 1.34 & 4.0
 & -0.71 &  9.3 & 2/3 & 0/3 &  5.0 & 0/3 & 0/3 \\
\midrule

\multirow{4}{*}{\textbf{ISOLET}}
 & ANN      & 93.46 & 2.65 & 3.0
 &  0.58 & 50.0 & 3/3 & 3/3 & 26.7 & 3/3 & 3/3 \\
 & AdaptHD  & 88.82 & 6.01 & 3.7
 & \textbf{10.37} & \textbf{19.7} & 3/3 & 3/3 & \textbf{50.0} & 3/3 & 3/3 \\
 & OnlineHD & 86.59 & 6.69 & 2.7
 & \textbf{15.65} & \textbf{27.7} & 3/3 & 3/3 & \textbf{50.0} & 3/3 & 3/3 \\
 & Vanilla  & 88.37 & 6.82 & 2.7
 & 42.21 &  8.0 & 2/3 & 2/3 & 50.0 & 3/3 & 3/3 \\
\midrule

\multirow{4}{*}{\textbf{UCI-HAR}}
 & ANN      & 90.82 & 4.68 & 7.7
 & -0.34 & 50.0 & 3/3 & 3/3 & 31.7 & 3/3 & 3/3 \\
 & AdaptHD  & 92.67 & 3.65 & 7.7
 &  1.19 & 28.0 & 3/3 & 2/3 & 50.0 & 3/3 & 3/3 \\
 & OnlineHD & 91.75 & 1.97 & 9.3
 & \textbf{25.20} & \textbf{14.0} & \textbf{1/3} & \textbf{0/3} & \textbf{50.0} & \textbf{3/3} & \textbf{3/3} \\
 & Vanilla  & 86.08 & 5.62 & 7.7
 & 17.78 & 10.0 & 1/3 & 1/3 & 21.7 & 2/3 & 2/3 \\
\bottomrule
\end{tabular}
}
\endgroup
\end{table*}

%% file: Tables/fpga.tex
\begin{table*}[ht!]
\centering
\caption{Watermarked AdaptHD under FPGA undervolting and overclocking (ZCU104). Accuracies (\%) and trigger matches (WM, out of $50$) average three hardware runs per setting. $E$ denotes mean core-supply energy per query.}

\label{tab:fpga}

\resizebox{0.7\textwidth}{!}{%

\begin{tabular}{ccccccccccc}
\toprule
\multicolumn{2}{c}{\textbf{Setting}} & & \multicolumn{2}{c}{\textbf{MNIST}} & \multicolumn{2}{c}{\textbf{FashionMNIST}} & \multicolumn{2}{c}{\textbf{ISOLET}} & \multicolumn{2}{c}{\textbf{UCI-HAR}} \\
\cmidrule(lr){1-2} \cmidrule(lr){4-5} \cmidrule(lr){6-7} \cmidrule(lr){8-9} \cmidrule(lr){10-11}
$V$ & $f$ & \textbf{E ($\mu$J)} & \textbf{Acc.} & \textbf{WM} & \textbf{Acc.} & \textbf{WM} & \textbf{Acc.} & \textbf{WM} & \textbf{Acc.} & \textbf{WM} \\
\midrule
0.852 V & 187.5 MHz & 204.4 & 95.00 & 50.0$\checkmark$ & 84.15 & 48.0$\checkmark$ & 93.26 & 50.0$\checkmark$ & 93.52 & 50.0$\checkmark$ \\
0.652 V & 187.5 MHz & 112.0 & 95.00 & 50.0$\checkmark$ & 84.15 & 48.0$\checkmark$ & 93.26 & 50.0$\checkmark$ & 93.52 & 50.0$\checkmark$ \\
0.645 V & 230.0 MHz & 108.2  & 95.00 & 50.0$\checkmark$ & 84.15 & 48.0$\checkmark$ & 93.26 & 50.0$\checkmark$ & 93.52 & 50.0$\checkmark$ \\
0.652 V & 250.0 MHz & 117.8 & 83.00 & 36.7$\checkmark$ & 74.58 & 31.0$\checkmark$ & 90.49 & 44.0$\checkmark$ & 89.38 & 40.3$\checkmark$ \\
0.645 V & 250.0 MHz & 116.2 & 73.00 & 31.0$\checkmark$ & 66.40 & 25.3$\checkmark$ & 88.75 & 41.3$\checkmark$ & 87.47 & 33.0$\checkmark$ \\
\textbf{0.652 V} & \textbf{287.5 MHz} & \textbf{106.5} & \textbf{21.36} & \textbf{6.3}$\times$      & \textbf{19.73} & \textbf{5.7}$\times$      & \multicolumn{2}{c}{\textbf{Hung}} & \multicolumn{2}{c}{\textbf{Hung}} \\
\bottomrule
\end{tabular}
}
\end{table*}

% Board energy:
% 841.4 
% 761.3 
% 712.1 
% 708.5 
% 704.4 
% 642.8 

%% file: Sections/hardware.tex
\section{Hardware Deployment and Fault Resilience}

Because HDC classifiers are frequently deployed in resource-constrained edge accelerators, where aggressive voltage scaling is used to reduce power consumption~\cite{zhang2022energy}, or an adversary can exploit voltage and clock for fault injection (glitching)~\cite{krautter2018fpgahammer,omidi2026faults}, we evaluate whether watermarked AdaptHD models remain verifiable under physical hardware faults. We deploy the watermarked $D = 10,000$ AdaptHD classifiers ($K = 50$, $\rho = 1\%$) onto a custom integer accelerator on an AMD/Xilinx Zynq UltraScale+ (ZCU104) FPGA executing both encoding and prototype similarity scoring in hardware. Starting from the nominal hardware operating point ($\text{VCCINT} = 0.852\text{ V}$, $f_{\text{clk}} = 187.5\text{ MHz}$), we induce timing-violation faults by undervolting the core supply rail down to $0.645\text{ V}$ while overclocking the accelerator from $187.5\text{ MHz}$ to $287.5\text{ MHz}$ across three runs per setting.

Table~\ref{tab:fpga} reports the mean core-supply energy per query ($E$), classification accuracy, and watermark trigger matches across representative voltage and frequency regimes. Lowering the supply voltage from the nominal $0.852\text{ V}$ baseline to $0.652\text{ V}$ at $187.5\text{ MHz}$ reduces mean core energy per query by $45.2\%$ ($204.4\text{ }\mu\text{J}$ to $112.0\text{ }\mu\text{J}$) while preserving identical classification accuracy and trigger matches ($48.0$ to $50.0$) across all four datasets. Pushing the accelerator further to $0.645\text{ V}$ and $230.0\text{ MHz}$ reduces energy to $108.2\text{ }\mu\text{J}$ per query ($47.1\%$ below nominal) while still maintaining unchanged task accuracy and watermark match counts.

When the clock frequency is raised to $250.0\text{ MHz}$, timing violations begin to degrade both task predictions and trigger agreement, yet the embedded watermark remains verifiable at $p < 10^{-3}$ in every run across all four datasets. Specifically, at $0.652\text{ V}$ and $250.0\text{ MHz}$, test accuracy drops by $2.77\text{ to }12.00\text{ pp}$ relative to the nominal baseline while mean trigger matches remain between $31.0$ and $44.0$ out of $50$. Further lowering the voltage to $0.645\text{ V}$ at $250.0\text{ MHz}$ increases the accuracy drop to $4.51\text{–}22.00\text{ pp}$ and reduces mean trigger matches to $25.3\text{–}41.3$, which still remains well above the $p < 10^{-3}$ verification thresholds ($m_{\min} \in \{8, 14, 18\}$) on every dataset. Watermark verification fails at $p < 10^{-3}$ only when overclocking reaches $287.5\text{ MHz}$ ($0.652\text{ V}$), where severe timing failures cause execution to hang on ISOLET and UCI-HAR and collapse classification accuracy to $21.36\%$ on MNIST ($6.3$ matches) and $19.73\%$ on FashionMNIST ($5.7$ matches).

These results show that because the watermark is distributed across the same high-dimensional class prototypes as the primary task representations, it shares the hardware fault tolerance of HDC; undervolting and overclocking either preserve watermark verification at $p < 10^{-3}$ or render the classifier unusable for its primary task.

%% file: Sections/conclusion.tex
\section{Conclusion}

We presented the first systematic investigation of black-box watermarking in HDC across software attacks and physical FPGA deployment. Error-driven learners reliably embed verifiable watermarks at 1\% exposure with negligible accuracy loss, whereas single-pass Vanilla HDC is bottlenecked by superposition cross-talk. Watermarks in error-driven HDC survive native continued training, and FPGA undervolting and overclocking experiments confirm that verification endures substantial hardware timing errors as long as the classifier remains functional. Under competing watermark overwriting, consolidating task and trigger representations into shared prototypes makes HDC more prone to erosion than distributed ANNs, yet an adversary still cannot erase the owner's signature without collapsing primary task accuracy, leaving survivable dual-verification states resolvable via cryptographic timestamping. In contrast, prediction-only model extraction produces high-fidelity surrogates that do not inherit OOD triggers. These findings motivate HDC-specific watermarking methods that account for its distinctive representation and learning dynamics to strengthen ownership verification.